\documentclass[aps,pra,reprint,showpacs,amsmath,superscriptaddress]{revtex4-1}
\usepackage{amssymb}
\usepackage{mathrsfs}
\usepackage{graphicx}
\usepackage{epstopdf}
\usepackage{float}
\usepackage{txfonts}
\usepackage[normalem]{ulem}
\usepackage{color}
\usepackage{CJK}
\usepackage{appendix}
\usepackage{bm}

\begin{document}
\title{Topological invariants and phase diagrams for one-dimensional
two-band non-Hermitian systems without chiral symmetry}

\author{Hui Jiang}
\affiliation{Beijing National Laboratory for Condensed Matter Physics, Institute of Physics, Chinese Academy of Sciences, Beijing 100190, China}
\affiliation{School of Physical Sciences, University of Chinese Academy of Sciences, Beijing 100049, China}
\author{Chao Yang}
\affiliation{Beijing National Laboratory for Condensed Matter Physics, Institute of Physics, Chinese Academy of Sciences, Beijing 100190, China}
\affiliation{School of Physical Sciences, University of Chinese Academy of Sciences, Beijing 100049, China}
\author{Shu Chen}
\email{schen@iphy.ac.cn}
\affiliation{Beijing National Laboratory for Condensed Matter Physics, Institute of Physics, Chinese Academy of Sciences, Beijing 100190, China}
\affiliation{School of Physical Sciences, University of Chinese Academy of Sciences, Beijing 100049, China}
\affiliation{Collaborative Innovation Center of Quantum Matter, Beijing, China}

\begin{abstract}
  We study topological properties of one-dimensional non-Hermitian systems without chiral symmetry and give phase diagrams characterized by topological invariants $\nu_E$ and $\nu_{tot}$, associated with complex energy vorticity and summation of Berry phases of complex bands, respectively.  In the absence of chiral symmetry, we find that the phase diagram determined by $\nu_E$ is different from $\nu_{tot}$.  While the transition between phases with different $\nu_{E}$ is closely related to the band-touching point, the transition between different $\nu_{tot}$ is irrelevant to the band-touching condition.  We give an interpretation for the discrepancy from the geometrical view by analyzing the relation of topological invariants with the winding numbers associated with exception points of the system.
  We then generalize the fidelity approach to study the phase transition in the non-Hermitian system and find that transition between phases with different $\nu_{tot}$ can be well characterized by an abrupt change of fidelity and fidelity susceptibility around the transition point.
\end{abstract}

\maketitle
\section{Introduction}
Topological phases of matter have been one of the most intriguing research subjects in condensed matter physics. Recently topological phases in non-Hermitian system have attracted great attention \cite{Bender1998,Bender,Hu2011,Esaki2011,Liang2013,Malzard2015,ZhuBG,Lee2016,Lee2,Leykam2017,Xu2017,SZ1,SZ2,Lieu2,ZengQB,Gonzsalez2017,Xiong2017,shen1,Yin,shen2,Ueda1,Ueda2,Rudner9,Rudner15,Gong10,Yuce15,Lieu1,Yuce2018,Jin2017,Torres1,Torres2,FuLiang1708,FuLiang1802,Ashida,Kim,Longhi,ZhangXD} partially motivated by the experimental progress on optical and optomechanical systems with gain and loss, which can be implemented in a controllable manner and effectively described by non-Hermitian systems \cite{Ruter2010,Peng2014,Feng2014,Konotop2016,Xiao2017,Weimann2017,Menke,Klett}. Recent studies have unveiled that the topological properties of non-Hermitian systems may exhibit quite different behaviors from  Hermitian systems, associated with some peculiar properties of the non-Hermitian Hamiltonian, e.g., biorthonormal eigenvectors, complex eigenvalues, the existence of exceptional points (EPs) and unusual bulk-edge correspondence  \cite{Heiss,Dembowski,Berry,Rotter,Hu2017,Hassan2017,Bergholtz,WangZhong1,WangZhong2,JingHui}.  Although non-Hermiticity brings some challenges for carrying out topological classification and properly defining topological invariants on biorthonormal eigenvectors \cite{Ueda1,Lieu2,WangZhong1}, the non$-$Hermitian system with novel qualities has opened up new frontiers for exploring rich topological phenomena.\\

\indent It is well known that symmetry and dimension play an important role in the study of topological properties \cite{Ueda1,Lieu2,Ueda2,Chiu16}.  For one-dimensional (1D) topological systems with chiral symmetry, the topological properties of the Hermitian systems can be characterized by a winding number $\nu_s$, which is closely related to the Berry phase across the Brillouin zone (Zak phase) of systems \cite{Zak,Yin,Li2015}. For the non-Hermitian system with chiral symmetry, one can generalize the definition of winding number $\nu_s$ as a topological invariant. Furthermore, due to
the eigenvalue being complex,
we need define another topological winding number $\nu_E$, describing the vorticity  of energy eigenvalues \cite{Leykam2017,shen1}. The phase diagram of
the non-Hermitian system with chiral symmetry can be well characterized by $\nu_E$ and $\nu_s$, which can take half integers. In a recent work \cite{Yin}, it was demonstrated that both  $\nu_E$ and $\nu_s$ are related to two winding numbers $\nu_1$ and $\nu_2$ which represent the times of trajectory of Hermitian part of the momentum-dependent Hamiltonian encircling the EPs.

 \indent  In this work, we study 1D non-Hermitian systems without chiral symmetry, which are found to exhibit quite different behaviors from their counterparts with chiral symmetry. In the absence of chiral symmetry, while $\nu_E$ remains to be a topological invariant, the Berry phase for each band is not quantized and the corresponding $\nu_s$ is no longer a topological invariant. Nevertheless, the summation of  $\nu_{s}$ for all the bands, denoted by $\nu_{tot}$, is still quantized and can be taken as  topological invariant \cite{Liang2013}.  By studying a concrete two-band non-Hermitian model, we find that the phase diagram determined by the topological invariant $\nu_E$ is different from that characterized by $\nu_{tot}$. While the phase boundaries of phase diagram characterized by $\nu_E$ correspond to the band-touching points of the non-chiral system, no band touching occurs at the phase boundaries of $\nu_{tot}$. This is in sharp contrast to the chiral non-Hermitian system, for which the phase boundaries between phases with different $\nu_{s}$ also correspond to the band-touching points. To understand the discrepancy of phase diagrams of the non-chiral systems, we further unveil the geometrical meaning of the topological invariants $\nu_E$ and $\nu_{tot}$.  Similar to the chiral non-Hermitian system, we find that $\nu_E$ is related to the winding numbers $\nu_1$ and $\nu_2$ which count the times  of trajectory of the Hermitian part of the Hamiltonian encircling the EPs of the non-chiral Hamiltonian. However, $\nu_{tot}$ is related to different winding numbers  $\nu'_1$ and $\nu'_2$ associated with EPs of a Hamiltonian in the absence of the term breaking the chiral symmetry.

 For the Hermitian system, besides the general Landau criteria for quantum phase transitions (QPTs), fidelity approach provides an alternative way to identify the QPT  from the perspective of wave functions \cite{zhu2006,Zan2006,Gu2007,chen2008,Ma2010}. Generally one may expect that the fidelity of ground state shows an abrupt change in the vicinity of the phase transition point of the system as a consequence of the dramatic change of the structure of the ground state. So far the studies of fidelity as a measure of QPTs are focused on Hermitian systems, for which either the Landua's energy criteria or the fidelity approach gives a consistent phase diagram. In this work, we shall generalize the fidelity approach to study phase transition in non-Hermitian systems. To our surprising, we find that both the fidelity and fidelity susceptibility exhibit obvious changes in the vicinity of phase boundaries of phases characterized by $\nu_{tot}$, instead of $\nu_E$. This suggests that the phase transition between phases with different $\nu_{tot}$ can be determined by the fidelity approach, whereas the transition between different $\nu_{E}$ is closely related to the band-touching (gap-closing) condition and can be determined by the generalized Landau's criteria. \\

\indent  The paper is organized as follows. In section II, we first give a general framework to expound the basic characteristics of two-band non-Hermitian system. In section III, we introduce a non-Hermitian model without chiral symmetry and analyze the spectrum of the system. We also calculate the topological invariant $\nu_E$ and give the phase diagram characterized by $\nu_E$. In section IV, we calculate the other topological invariant $\nu_{tot}$, associated with the Berry phase, and the phase diagram characterized by $\nu_{tot}$. We find discrepancy of phase diagrams characterized by $\nu_E$ and $\nu_{tot}$, and unveil that the two topological invariants are related to different winding numbers associated with the EPs of the Hamiltonian with and without chiral symmetry. We also analyse the effect of a hidden pseudo-inversion symmetry on the topological property of eigenstate. Then, we calculate the fidelity of a given eigenstate and the corresponding fidelity susceptibility to identify the phase transition characterized by $\nu_{tot}$.
A summary is given in the last section.

\section{Topological invariants of 1D two-band non-Hermitian systems}
In general, a two-band non-Hermitian system can be described by \\
 \begin{equation}\label{1}
  \mathcal{H}(k)= \emph{\textbf{h}}(k) \cdot {\boldsymbol \sigma} = \emph{\textbf{n}}(\emph{k}) \cdot  {\boldsymbol \sigma} +\mathrm{i} \bm{\gamma} (\emph{k})\cdot  {\boldsymbol \sigma},
 \end{equation}
 where $\emph{\textbf{h}}(k)$,  $\emph{\textbf{n}}(\emph{k})$ and $\bm{\gamma} (\emph{k})$  may include three components $x, y ,z$ and $\sigma_{x,y,z}$ is the Pauli matrix. In general, the non$-$Hemitian system can be divided into the summation of  Hermitian and non-Hermitian part: $\emph{\textbf{h}}(k)= \emph{\textbf{n}}(\emph{k}) + \mathrm{i}  \bm{\gamma} (\emph{k})$  with $\emph{\textbf{n}}(\emph{k})$ and $\bm{\gamma} (\emph{k})$ being real functions of $k$. The energy square of non$-$Hemitian Hamiltonian is: $E^2= |\emph{\textbf{n}}|^2-|\bm{\gamma} |^2+2 \mathrm{i}  \emph{\textbf{n}} \cdot {\bm{\gamma}}:=E_1^2=E_2^2$ ($E_1=-E_2$). It is clear that the two bands touch at zero when  $\emph{\textbf{n}}(k)$$\perp$$ {\bm{ \gamma}}(k) $ and $|\emph{\textbf{n}}(k)|=|{\bm {\gamma}}(k)|$.\\

 \indent The eigenvalue $E_{1,2}$ is smoothly continuous with $k$. Since the eigenvalue is generally complex, we can represent it as  $E_1$=$|E|e^{i\theta_k}$=$-E_2$ with $\theta_k$  the angle of eigenvalue. As $k$ goes across the Brillouin zone (BZ), we can always define the winding number of energy $\nu_{E}$ as \cite{Leykam2017,shen1}
 \begin{equation}\label{nuE}
   \nu_E=\frac{1}{2\pi}\oint dk\partial_k \mathrm{Arg}(\triangle E)=\frac{1}{2\pi}\oint dk\partial_k \mathrm{Arg}(E_1-E_2).
 \end{equation}
For the Hermitian system, $\nu_E$ is always zero as $\theta_k$ takes either $0$ or $\pi$.  See appendix A for the detailed calculation of $\nu_E$.

 \indent On the other hand, the eigenstates of non-Hermitian Hamiltonian (Eq.(\ref{1})) satisfy $\mathcal{H}(\emph{k})|\psi^{R}_{1,2}\rangle=E_{1,2}|\psi^{R}_{1,2}\rangle$, and $|\psi^{R}_{1,2}\rangle$ do not form an orthogonal basis. In order to describe non-Hermitian properties, we need also consider the eigenstates of $\mathcal{H}^{\dagger}$, $\mathcal{H}^{\dagger}(\emph{k})|\psi^{L}_{1,2}\rangle=E^{*}_{1,2}|\psi^{L}_{1,2}\rangle$, which together with $|\psi^{R}_{1,2}\rangle$ form biorthogonal vectors and fulfill $\langle\psi^{L}_{i}|\psi^{R}_{j}\rangle=\delta_{i，j}$ by properly choosing the normalization $\langle\psi^{L}_{1,2}|\psi^{R}_{1,2}\rangle$.
For simplicity, we choose
\[
|\psi^{R}_{1,2}\rangle=\frac{1}{\sqrt{2E_{1,2}(E_{1,2}-h_z)}}\left(
                           \begin{array}{cc}
                             h_x-ih_y &E_{1,2}-h_z \\
                           \end{array}
                         \right)^{T},
\]
where the superscript $T$ is transpose operation, and
\[
\langle\psi^{L}_{1,2}|=\frac{1}{\sqrt{2E_{1,2}(E_{1,2}-h_z)}}\left(
                                                                                 \begin{array}{cc}
                                                                                   h_x+ih_y &E_{1,2}-h_z \\
                                                                                 \end{array}
                                                                               \right).
\]
Similar to the definition of winding number related to the Berry phase of eigenstate in Hermitian system, one can generalize the definition $\nu_{s}$ directly to the non-Hermitian system \cite{ZhuBG,Yin,Lieu1}, which can be written as
\begin{equation}
  \nu_{s,\alpha}=\frac{1}{\pi}\oint dk \langle\psi^{L}_{\alpha}|\mathrm{i}\partial_{k}|\psi^{R}_{\alpha}\rangle,
\end{equation}
where $\alpha=1, 2$ indicate the band labels.
Substituting the concrete forms  of $|\psi^{R}_{-}\rangle$ and $\langle\psi^{L}_{-}|$ into the above equation, after some simplifications, we can represent $\nu_{s}$ as (Appendix.B):
\begin{equation}\label{23}
  \nu_{s,\alpha}=\frac{1}{2\pi}\oint dk\frac{h_x\partial_{k}h_y-h_y\partial_{k}h_x}{E_\alpha(E_\alpha-h_z)},
\end{equation}
where $E_1$ and $E_2$ are eigenvalues of the non-Hermitian Hamiltonian.

For the case with chiral symmetry, it has been shown that both $\nu_E$ and $\nu_{s, \alpha}$ can only take some half-integers. In a recent work, it has been demonstrated  $\nu_E$ and $\nu_{s ,\alpha}$  are related to the winding numbers $\nu_1$ and $\nu_2$ of trajectory of the Hermitian part around two different EPs, respectively \cite{Yin}, and thus explain why they are topological invariant with half-integers. The phase diagrams can be determined by different values of either $\nu_E$ or $\nu_{s,\alpha}$, or equivalently $\nu_1$ and $\nu_2$. For the general case without chiral symmetry, $\nu_E$ remains to be a topological invariant, however, $\nu_{s ,\alpha}$ is generally a complex number which is not quantized,  suggesting that   $\nu_{s ,\alpha}$ is no longer a topological invariant. Nevertheless,
\[
\nu_{tot}=\nu_{s,1} + \nu_{s,2}
\]
has been demonstrated to be a topological invariant, which takes integers \cite{Liang2013}. As shall be discussed in detain in the following section, we find that phase boundaries of the phase diagram determined by $\nu_E$ is consistent with the band touching curves determined by $E_1 = E_2 = 0$. On the other side, we can also get a phase diagram determined by topological invariant $\nu_{tot}$, which displays obviously different phase boundaries from phase boundaries determined by $\nu_E$. To understand this discrepancy, we further analyze geometrical origins of
$\nu_E$ and $\nu_{tot}$, associated to the Hamiltonian (\ref{Hk}). While $\nu_E$ can be related to the winding numbers around to two EPs of the Hamiltonian  (\ref{Hk}) via $\nu_E = \pm \frac{1}{2} (\nu_2 - \nu_1)$, we find no relation of $\nu_{tot}$ with $\nu_1$ and $\nu_2$, instead we have $\nu_{tot}= \nu'_1 + \nu'_2$, where $\nu'_1$ and $\nu'_2$ are winding numbers around EPs of the Hamiltonian in the absence of the chemical potential term.

\section{Model and spectrum}

For simplicity, we consider a 1D non-Hermitian model by choosing the Su-Schrieffer-Heeger (SSH) model as the Hermitian part of the non-Hermitian Hamiltonian, and introduce an off-diagonal non-Hermitian part by taking different hopping amplitudes along the right and left hopping directions in the unit cell \cite{Yin,WangZhong1}. A diagonal non-Hermitian term is also introduced by alternatively adding imaginary chemical potential $\pm i \mu $ on the A/B$-$sublattice. Explicitly, the Hamiltonian is given by
\begin{equation}\label{hiu}
\begin{split}
H&=\sum_n (t+\delta)c^{\dagger}_{A,n}c_{B,n}+(t-\delta)c^{\dagger}_{B,n}c_{A,n}+t'c^{\dagger}_{A,n+1}c_{B,n}\\
&\quad \quad \quad \quad\quad +t'c^{\dagger}_{B,n }c_{A,n+1}+ i \mu  c^{\dagger}_{A,n}c_{A,n}- i \mu c^{\dagger}_{B,n}c_{B,n},
\end{split}
\end{equation}
with $t'=1$ as the unit of energy in the following discussion. Under the periodic boundary condition, we can make a Fourier transformation: $ c_{\alpha,n} =1/\sqrt{N}\sum_k e^{ikn}c_{\alpha,k}$
where $N$ is the number of the unit cells and $\alpha$ takes $A$ or $B$. Then the Hamiltonian can be written in the form of
\begin{equation}\label{56}
  H(\emph{k})=\sum_{k} \phi^{\dagger}_{\emph{k}}\mathcal{H}(k) \phi_{\emph{k}},
\end{equation}
where $\phi^{\dagger}_{{k}}=(c^{\dagger}_{A,k},c^{\dagger}_{B,k})$, and
\begin{equation}
\mathcal{H}(k) = \left(
      \begin{array}{cc}
        i \mu   & t+\delta+e^{-ik} \\
        t-\delta+e^{ik} &-i \mu \\
      \end{array}
    \right)
    = \emph{\textbf{n}}(\emph{k})\cdot  {\boldsymbol \sigma} +\mathrm{i}\delta \sigma_y+\mathrm{i} \mu  \sigma_z . \label{Hk}
\end{equation}
Here the Hermitian part is $\emph{\textbf{n}}(\emph{k})\cdot  {\boldsymbol \sigma}  = n_x(k) \sigma_x +  n_y(k) \sigma_y$ with $n_x=t+\cos k$ and $n_y= \sin k $. When $\mu=0$, the term of $\sigma_z$ vanishes and the model reduces to the chiral non-Hermitian SSH model which fulfills the chiral symmetry \cite{Yin}:
\[
\sigma_z \mathcal{H}(k) \sigma_z = - \mathcal{H}(k).
\]
The chiral symmetry is broken when $\mu \neq 0$.

\begin{figure}[htbp]
  \centering
  \includegraphics[width=3.0in]{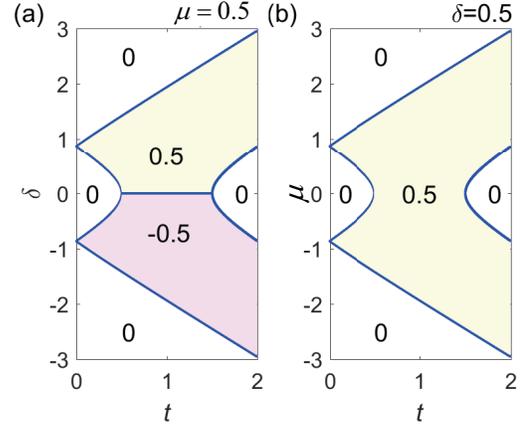}\\
  \caption{Phase diagram characterized by the winding number of energy  $\nu_E$.  (a) $t$ versus $\delta$ by fixing $\mu=0.5$, and (b) $t$ versus $\mu$ by fixing $\delta=0.5$. The light yellow shallow represents the winding number
  of energy $\nu_E=0.5$ and the light pink  shallow represents  $\nu_E=-0.5$, while other regimes are  $\nu_E=0$. The phase transition is accompanied by the band touching (close of band gap).} \label{a}
\end{figure}
\begin{figure*}[htbp]
  \centering
  \includegraphics[width=6.8in]{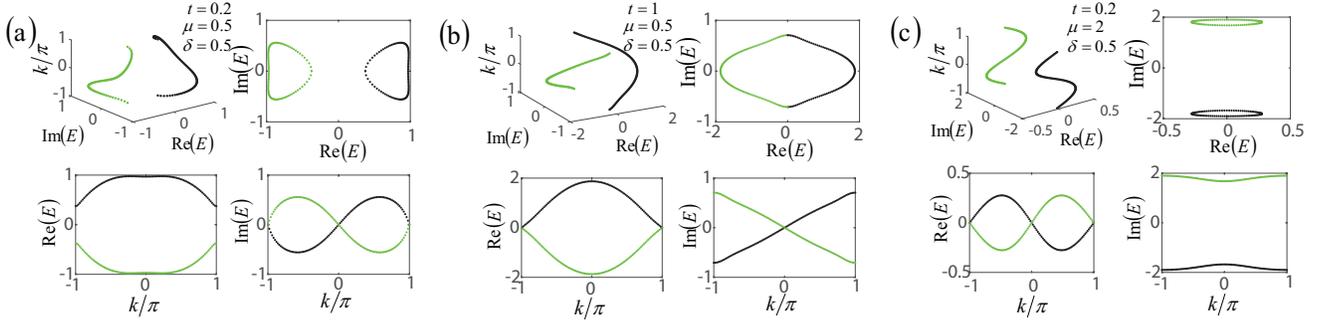}\\
  \caption{Energy distribution in different parameter regions. The green curve and the black one represent $E_1$ and $E_2$, respectively. The detailed parameters are shown in the figure. (a) and (c) correspond to  $\nu_E=0$,  and (b) corresponds to $\nu_E=0.5$.}\label{b}
\end{figure*}

From Eq.(\ref{Hk}), it is straightforward to get the square of eigenvalues given by
\begin{equation*}
\begin{aligned}
  E^2(k)=t^2+1+2t\cos k-\delta^2 - \mu^2 -2i\delta\sin k ,
  \end{aligned}
\end{equation*}
which suggests the existence of two solutions $E_{1}$ and $E_{2}$ with $E_{2}=-E_{1}$.
The $i$-th band energy $E_{1,2}$ can be represented as $E_i(k)=|E_i(k)|e^{i\theta_i(k)}$ ($i=1,2$ ) where $\theta_2(k)=\theta_1(k)+\pi=\theta(k)+\pi$. Substituting $E_{1,2}$ into Eq.(\ref{nuE}), we can simplify $\nu_E$ to
\begin{eqnarray}
\begin{aligned}
  \nu_E &= \frac{1}{4}\sum_{i}\mathrm{sgn}(\delta)\mathrm{sgn}(\frac{\partial n_y}{\partial k }\mid_{K_i}) \\ &\qquad\qquad\cdot\mathrm{sgn}\left((n^2_x-\delta^2-\mu^2 )\mid_{K_i}\right),
  \end{aligned}
\end{eqnarray}
where $k=K_i$ is the $i$-th solution of $n_y=0$, which gives $k=0$ and $\pi$.
In Fig.1 we show the phase diagram of the model (\ref{hiu}) with different phases characterized by different $\nu_E$.  In Fig.1(a), the phase diagram is plotted for $t$ versus $\delta$ by fixing  $\mu=0.5$, and Fig.1(b) is for $t$ versus $\mu$ by fixing  $\delta=0.5$. We find that the phase boundaries can be determined by $ \delta^2 + \mu^2 = (t \pm 1)^2$, which is consistent with the band-touching (gap-closing) condition $E_1(k)=E_2(k)=0$, i.e., the two bands touch together at the phase boundaries.

It is shown that in some regions of the phase diagram $\nu_E$ takes the half integer $\pm 1/2$, which suggests the definition Eq.(\ref{nuE}) is not a true winding number in the geometrical meaning. The reason behind this is that in this region the complex eigenvalue $E_{1}(k)$ or $E_{2}(k)$ does not form a close curve when $k$ goes around the BZ.  To see it clear, we show $E_i(k)$ versus $k$ in Fig.$\ref{b}$, in which $E_i(k)$ changes continuously and smoothly with $k$. 
As shown in Fig.$\ref{b}$ (b), neither $E_1$ nor $E_2$ form a close curve as $k$ changes from $-\pi$ to $\pi$, instead they switch each other with $E_{1}(\pi) =  E_{2}(-\pi)$ and $E_{2}(\pi) =  E_{1}(-\pi)$, in contrast with the phase regimes with $\nu_E=0$ corresponding to Fig.$\ref{b}$ (a) and (c), where both $E_{1,2}(k)$ forms a close curve and we have $E_{i}(\pi) =  E_{i}(-\pi)$.

Furthermore, we demonstrate that the definition Eq.(\ref{nuE}) is equivalent to half of the difference of two winding numbers, i.e.,
\begin{equation}
\nu_E = \frac{1}{2} \mathrm{sgn}(\delta) (\nu_2 - \nu_1),
\end{equation}
where $\nu_{1,2} =\frac{1}{2 \pi} \oint dk \nabla_k  \phi_{1,2}$ with $\phi_{1,2}$ defined by
\[
\tan \phi_1=\frac{n_y}{n_x+\sqrt{\mu^2+\delta^2}}, ~~~~
\tan \phi_2=\frac{n_y}{n_x-\sqrt{\mu^2+\delta^2}}.
\]
It is clear that $\nu_{1}$ and $\nu_{2}$ represent the winding number of the closed curve formed by $(n_x(k), n_y(k))$ in the two-dimensional space surrounding the EPs $(-\sqrt{\mu^2+\delta^2},0)$ and $(\sqrt{\mu^2+\delta^2},0)$, respectively.

\section{Topological properties of eigenvectors}
\subsection{Topological invariant of eigenvectors}
\begin{figure}[htbp]
  \centering
  \includegraphics[width=3.0in]{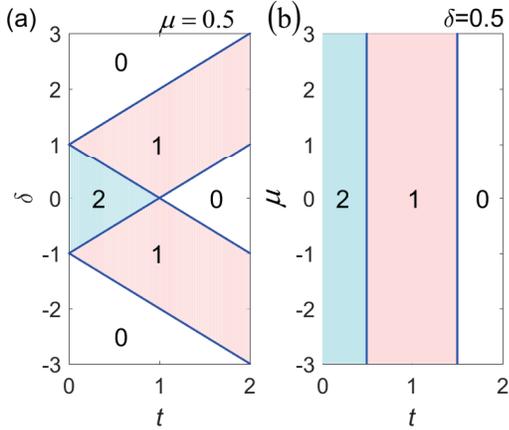}\\
  \caption{Phase diagram characterized by topological invariant $\nu_{tot}$.  (a) $t$ versus $\delta$ by fixing $\mu=0.5$ and (b) $t$ versus $\mu$ by fixing $\delta=0.5$. The number in different color areas represents the topological invariant $\nu_{tot}=\nu_{s,1}+\nu_{s,2}$.} \label{phasediagram2}
\end{figure}

By using the expression of Eq.(\ref{23}) and substituting it into $\nu_{tot}= \nu_{s,1}+\nu_{s,2}$, we get
\begin{equation*}
  \nu_{tot} = \frac{1}{2\pi}\oint dk \left[ \frac{h_x\partial_{k}h_y-h_y\partial_{k}h_x}{E_1(E_1-h_z)}+ \frac{h_x\partial_{k}h_y-h_y\partial_{k}h_x}{E_2(E_2-h_z)} \right]. \\
\end{equation*}
With the help of the relation $E_2=-E_1$, the above equation can be rewritten as
\begin{eqnarray*}
  \nu_{tot} &=& \frac{1}{2\pi}\oint dk \left[ \frac{h_x\partial_{k}h_y-h_y\partial_{k}h_x}{E_1(E_1-h_z)}+ \frac{h_x\partial_{k}h_y-h_y\partial_{k}h_x}{E_1(E_1+h_z)} \right], \\
  &=& \frac{1}{\pi}\oint dk \frac{h_x\partial_{k}h_y-h_y\partial_{k}h_x}{E_1^2-h_z^2}.
\end{eqnarray*}
Since $E_1^2=h_x^2+h_y^2 + h_z^2$, we can get
\begin{equation}\label{31}
  \nu_{tot}= \frac{1}{\pi}\oint dk \frac{h_x\partial_{k}h_y-h_y\partial_{k}h_x}{h_x^2+h_y^2},
\end{equation}
where $h_x=n_x=t+\cos k$ and $h_y=n_y+ i \gamma_y = \sin k + i \delta$. We notice that $\nu_{tot}$ is independent of $h_z$, although its definition  is related to the eigenvectors of $\mathcal{H}(k)$.\\

In Fig.\ref{phasediagram2}, we show the phase diagram characterized by different values of $\nu_{tot}$. In Fig.\ref{phasediagram2}(a), the phase diagram is plotted for $t$ versus $\delta$ by fixing a $\mu=0.5$, and Fig.\ref{phasediagram2}(b) is for $t$ versus $\mu$ by fixing a $\delta=0.5$. Fig.\ref{phasediagram2}(b) clearly indicates that the phase diagram is irrelevant to $\mu$ as the expression of $\nu_{tot}$ is independent of $h_z$. From the expression of Eq.(\ref{31}), we can see that the phase diagram shown in Fig.\ref{phasediagram2}(a) is identical to the phase diagram of the Hamiltonian in the absence of $h_z$ term, i.e., the non-Hermitian Hamiltonian with chiral symmetry given by
\begin{equation}
\mathcal{H}_{chiral}(k) = (t+\cos k) \sigma_x + (\sin k + i \delta) \sigma_y.
\end{equation}
The expression Eq.(\ref{31}) does not represent a winding number in the geometrical meaning as $h_y(k)$ is not a real function. Following the same derivation for the case with chiral symmetry \cite{Yin}, we can represent $\nu_{tot}$ as the summation of two true winding numbers
\begin{equation}
\nu_{tot} = \nu'_1 + \nu'_2 \label{nutot},
\end{equation}
where $\nu'_{1,2} =\frac{1}{2 \pi} \oint dk \nabla_k  \phi'_{1,2}$ with $\phi'_{1,2}$ defined by
\[
\tan \phi'_1=\frac{n_y}{n_x+ \delta}, ~~~~
\tan \phi'_2=\frac{n_y}{n_x- \delta}.
\]
It is clear that $\nu'_{1}$ and $\nu'_{2}$ represent the winding number of the closed curve formed by $(n_x(k), n_y(k))$ in the two-dimensional space surrounding two points $(-\delta ,0)$ and $(\delta,0)$, respectively. These two points are not EPs of the Hamiltonian (\ref{Hk}), instead they are EPs of $\mathcal{H}_{chiral}(k)$. Consequently, the phase boundary of the phase diagram determined by $\nu_{tot}$ is same with the band touching condition for the system described by $\mathcal{H}_{chiral}(k)$, but is different from the phase diagram determined by $\nu_{E}$.

\begin{figure}[htbp]
  \centering
  \includegraphics[width=3.0in]{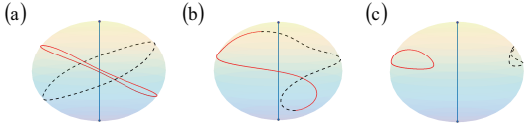}\\
  \caption{The unit sphere vector $\emph{\textbf{R}}(\emph{k})$ (red curve  with eigenvalue $E_1$ and black one with eigenvalue $E_2$). The parameter $t$  in (a), (b) and (c) takes $0$, $1$ and $2$, respectively, with other parameters $\mu=0.5$ and $\delta=0.5$. The blue line connects the north and south poles. }\label{i}
\end{figure}

\indent Alternatively, we can also understand the geometrical meaning of the topological invariant $\nu_{tot}$ from trajectories of eigenvectors by projecting the eigenvectors onto a 2D unit spherical surface. In general, the right-eigenvector can be parameterized as
\begin{equation}\label{k1}
  |\psi_R(\alpha_k,\beta_k)\rangle=\left(
                                     \begin{array}{c}
                                       \cos \frac{\beta_k}{2}\\
                                       e^{i\alpha_k}\sin \frac{\beta_k}{2}\\
                                     \end{array}
                                   \right),
\end{equation}
For each eigenvector corresponding to $E_1$ or $E_2$, we may calculate the sphere vector defined as $\emph{\textbf{R}}(\emph{k})$=$(\cos\alpha_k\sin\beta_k,\sin\alpha_k\sin\beta_k,\cos\beta_k)$, where $\alpha_k$ and $\beta_k$ correspond to the azimuthal and polar angles of $\emph{\textbf{R}}(\emph{k})$, respectively. In Fig.{\ref{i}}, we plot the evolution of two eigenvectors on the Bloch sphere across the Brillouin zone. Their trajectories form separately two closed curves as shown in Fig.{\ref{i}} (a) and (c),  or form together a close curve in Fig.{\ref{i}} (b).  The topological invariant $\nu_{tot}$ can be viewed as a winding number which accounts times of the trajectories  passing  around the z-axis connecting north and south poles.

\begin{figure}[htbp]
  \centering
  \includegraphics[width=3.0in]{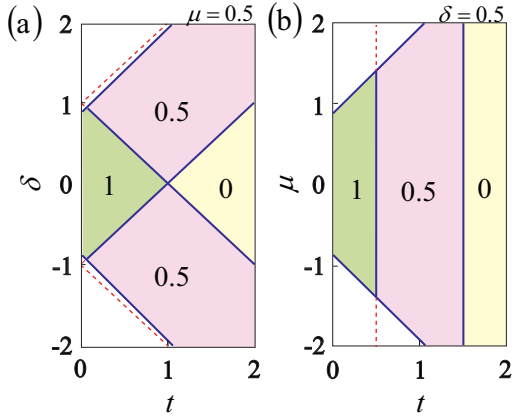}\\
  \caption{Phase diagram characterized by the real part of Berry phase $\nu_{s,1}$.  (a) $t$ versus $\delta$ by fixing $\mu=0.5$, and (b) $t$ versus $\mu$ by fixing $\delta=0.5$. The number in different color areas represents the quantized Re($\nu_{s,1}$), and in the regions without the number  Re($\nu_{s,1}$) is not quantized. The blue solid curve represents the phase boundary of phase diagram characterized by Re($\nu_{s,1}$), and the red dashed is corresponding to the phase boundary of $\nu_{tot}$.}\label{phasediagram3}
\end{figure}

\indent
Generally speaking, $\nu_{s,1}$ is not quantized for a system without the chiral symmetry. However, for the model described by Eq.(\ref{Hk}), the Hamiltonian satisfies a pseudo-inversion symmetry:
\begin{equation}
\sigma_x \mathcal{H}(k)\sigma_x= \mathcal{H}^{\dagger}(-k),
\end{equation}
and we find that the real part of $\nu_{s,1}$ is quantized in some parameter regions due to the existence of the pseudo-inversion symmetry.
Given that $\mathcal{H}(\emph{k})|\psi^{R}_{\alpha}(k)\rangle=E_{\alpha}(k)|\psi^{R}_{\alpha}(k)\rangle$, it follows
\[
\mathcal{H}^{\dagger}(-\emph{k})\sigma_x|\psi^{R}_{\alpha}(k)\rangle=E_{\alpha}(k)\sigma_x|\psi^{R}_{\alpha}(k)\rangle.
\]
Noticing that
$\mathcal{H}^{\dagger}(-k)|\psi^{L}_{\alpha}(-k)\rangle=E^*_{\alpha}(-k)|\psi^{L}_{\alpha}(-k)\rangle$, we have
$E_1(k)= E^{*}_1(-k)$ if the state fulfills $\sigma_x|\psi^{R}_{1}(k)\rangle = |\psi^{L}_{1}(-k)\rangle$ or
$E_1(k)= E^{*}_2(-k)$ if the state fulfills $\sigma_x|\psi^{R}_{1}(k)\rangle = |\psi^{L}_{2}(-k)\rangle$.
The difference between these two cases can be distinguished  by whether the real part of $\nu_{s,1}$  is quantized or not. The real part of $\nu_{s,1}$ is quantized in the case of $E_1(k)= E^{*}_1(-k)$, and $\nu_{s,1}$ is not quantized but real in the other case. In Fig.5, regions labeled by quantized number  $0$ ,$0.5$, $1$  correspond to the case of $E_1(k)= E^{*}_1(-k)$ with quantized real part of $\nu_{s,1}$. Regions without labeled numbers correspond to  the case of $E_1(k)= E^{*}_2(-k)$, for which $\nu_{s,1}$ is no longer quantized. The boundaries between these two cases can be determined by $E_{1,2}^{2}(k=0)=0$ (see appendix B for details).\\

When the chemical potential term $h_z$ is no longer imaginary, i.e, $h_z \equiv n_z + i \gamma_z = \eta + i \mu $ with nonzero $\eta$, the pseudo-inversion system is broken, and the real part of $\nu_{s,1/2}$ is not quantized. Nevertheless, $\nu_{tot}$ is always quantized and takes the same value no matter which form $h_z$ takes, i.e., the expression of  Eq.(\ref{nutot}) is irrelevant to the term of $h_z$.


\subsection{Detection of phase boundaries via fidelity approach}
\indent
We have demonstrated that the phase diagram determined by $\nu_{tot}$ displays quite different phase boundaries from the band-touching conditions. As  $\nu_{tot}$ reflects the global geometrical properties of wavefunctions, we apply the fidelity approach to detect the phase boundaries. The fidelity approach has been widely used to study the phase transitions in various quantum many-body systems \cite{zhu2006,Zan2006,Gu2007,chen2008,Ma2010}.
Given a Hamiltonian $H(\lambda)$, which depends on the driving parameter $\lambda$, the quantum fidelity is defined as the overlap between two eigen$-$states with only slightly different values of the external parameter and thus is a pure geometrical quantity.
For the non-Hermitian Hamiltonian studied in this work, the driving parameter $\lambda$ can be taken as $t$, $\delta$ or $\mu$. In terms of the eigenstates $|\psi_{R,n}(\lambda)\rangle$ of
$ H(\lambda)$, the Hamiltonian can be reformulated as $ H(\lambda) =
\sum_n E_n(\lambda)|\psi_{R,n}(\lambda)\rangle\langle\psi_{L,n}(\lambda)|$. Therefore, we can generalize the definition of the state fidelity to the non-Hermitian system, which is defined as the half sum of  the overlap between $|\psi_{R,1}(\lambda+\epsilon)\rangle $and $|\psi_{L,1}(\lambda)\rangle$ and the overlap between $|\psi_{L,1}(\lambda+\epsilon)\rangle $and $|\psi_{R,1}(\lambda)\rangle$, i.e.,
\begin{equation}\label{f1}
  F(\lambda,\epsilon)=\frac{1}{2}|\langle\psi_{L,1}(\lambda)|\psi_{R,1}(\lambda+\epsilon)\rangle+\langle\psi_{R,1}(\lambda)|\psi_{L,1}(\lambda+\epsilon)\rangle|,
\end{equation}
 where $|\psi_{R,1}(\lambda)\rangle$ is the wavefunction corresponding to the
parameter $\lambda$ with  eigenenergy $E_1$ and $\epsilon$ is a small quantity. It is obvious that the fidelity is dependent of $\epsilon$. The rate of change of fidelity is given by the second derivative of fidelity or fidelity susceptibility
\begin{equation}\label{f2}
  S(\lambda)= \lim_{\epsilon\rightarrow 0}\partial^2_{\epsilon}\mathrm{In} F(\lambda,\epsilon),
\end{equation}
which is independent of $\epsilon$.
We note that the first derivative of fidelity defined by Eq.(\ref{f1}) gives zero, which is consistent with the Hermitian system \cite{Zan2006,Gu2007,chen2008}.
\begin{figure}[htbp]
 \centering
  \includegraphics[width=3.5in]{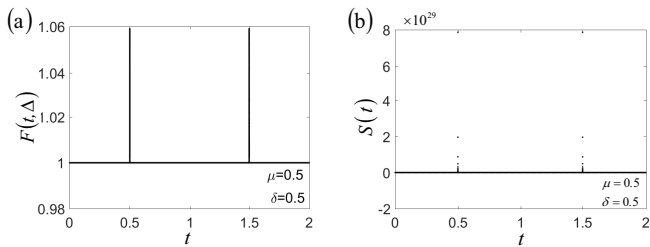}\\
  \caption{The fidelity (a) and fidelity susceptibility (b) as a function of $t$. Here we take $\lambda=t$, $\mu=0.5$ and $\delta=0.5$.}\label{d4}
\end{figure}

In Fig.\ref{d4}, we display the fidelity and fidelity susceptibility versus the driving parameter $t$, i.e., we take $\lambda=t$, by fixing $\delta=0.5$ and $\mu=0.5$.
It is shown that both the fidelity and fidelity susceptibility exhibit an abrupt jump in the vicinity of the transition points, which are consistent with the phase boundaries of the phase diagram determined by $\nu_{tot}$.
If we take the driving parameter as $\delta$ by fixing $t$ and $\mu$, similarly we find an abrupt jump of the fidelity and fidelity susceptibility in the vicinity of the transition points.
Our results demonstrate that the phase transition point determined by the fidelity approach is different from that obtained by using Landau's energy criterion, which gives the phase boundaries by the band crossing condition. For the Hermitian system, it has been demonstrated that the fidelity susceptibility and the second derivatives of ground energy play an equivalent role in identifying the quantum phase transition. However, for the non-Hermitian system, they play different roles and may give different phase boundaries when the chiral symmetry is broken. This also explains why the discrepancy of phase diagrams determined by $\nu_E$ and $\nu_{tot}$ may arise for the non-Hermitian system.

\section{Summary}
\indent In summary, we have studied 1D general non-Hermitian systems without chiral symmetry and found the existence of discrepancy between phase diagrams characterized by two independent topological invariants $\nu_E$ and $\nu_{tot}$, which are quantized for our studied systems.
While the phase boundaries between phases with different $\nu_E$ are determined by the band-touching condition, the phase boundaries between different $\nu_{tot}$ are irrelevant to the band touching of the non-chiral system.  The discrepancy of phase diagrams can be further clarified from the geometrical meaning the topological invariants $\nu_E$ and $\nu_{tot}$, which can be represented as $\nu_E=\pm (\nu_2-\nu_1)/2$ and $\nu_{tot}=\nu'_2+\nu'_1$, where $\nu_1$ and $\nu_2$ are winding numbers counting the times  of trajectory of the  Hermitian part of the Hamiltonian encircling two EPs of the non-chiral Hamiltonian, and $\nu'_1$ and $\nu'_2$ are winding numbers associated with two EPs of the Hamiltonian in the absence of the chiral-symmetry breaking term. The fact that the topological invariant $\nu_{tot}$ is independent of the chiral-symmetry breaking term suggests that the corresponding transition between different $\nu_{tot}$ is irrelevant to the band-touching points, instead it is equal to the winding number which counts times of trajectories of vectors by projecting the eigenstates onto 2D unit sphere passing around the z-axis connecting north and south poles.  Furthermore, we find the existence of a hidden pseudo-inversion symmetry and the real part of $\nu_{s,\alpha}$ is quantized when the  eigenvalues of the system satisfy $E_{1,2}(k)=E^{*}_{1,2}(-k)$.

 We then generalize the definition of fidelity and use the fidelity and fidelity susceptibility to identify the phase transition in the non-Hermitian system. Our results show that an abrupt change of fidelity and fidelity susceptibility occurs around transition points between phases with different $\nu_{tot}$, which suggests that the fidelity approach can witness topological phase transitions characterized by $\nu_{tot}$  accompanied with no gap closing in the non-Hermitian system.
Our work unveils that the non-Hermitian systems may exhibit some peculiar properties, which have no correspondence in the Hermitian systems and are worthy of further investigation. A question that remains open is to find physical observable quantities to detect the topological invariants in the non-Hermitian
models without chiral symmetry.

\begin{acknowledgments}
The work is supported by NSFC under Grants No. 11425419, the National Key Research and Development Program of China (2016YFA0300600 and 2016YFA0302104) and the Strategic Priority Research Program (B) of the Chinese Academy of Sciences  (No. XDB07020000).
\end{acknowledgments}
\appendix
\section{The winding of eigenenergy $\nu_E$}

The winding number of energies $\nu_E$ can be written as
\begin{equation*}
 \nu_E=\frac{1}{2\pi}\int\nabla_{\mathbf{k}}\mathrm{Arg}(\triangle E) d\mathbf{k},
\end{equation*}
where $\triangle E$ represents the difference of energies between any of the two bands. Generally speaking, a 2-band non-Hermitian system can be described by the Hamiltonian in Eq.(\ref{1}), with eigenvalues $E_{1,2}^2=|\emph{\textbf{h}}(k)|^2$ ($E_1=-E_2$). Hence the angle of $\triangle E$ is half of the angle of $E_{1,2}^2$, and as a result $\nu_E$ can be interpreted as the half of the winding number of $E_{1,2}^2$ in the complex plane around the origin. In Hermitian systems, the energy $E_{1,2}$ is real and $\nu_E$ is always zero.

Similar to Ref.\cite{zliu2018}, the winding number of $\nu_E$ can be written as
\begin{equation}
  \nu_E = \frac{1}{4}\sum_{i}(\mathrm{sgn}(\frac{\partial \mathrm{Im}(E_{1,2}^2)}{\partial \mathbf{k} }\mid_{\mathbf{k}=\mathbf{K}_i})\cdot \mathrm{sgn}(\mathrm{Re}(E_{1,2}^2)(\mathbf{K}_i)),
\end{equation}
with $\mathbf{K_i}$ being the $i-$th solution of $\mathrm{Im}(E_{1,2}^2)=0$. For the Hamiltonian described by Eq.(\ref{Hk}), the eigenvalues satisfy $E^2_{1,2}=t^2+1+2t\cos k-\delta^2-\mu^2 +2i\delta\sin k$. It's easy to get simplified form of $\nu_E$,
\begin{equation}
\begin{split}
  \nu_E = &\frac{1}{4}\sum_{i}\mathrm{sgn}(\delta)\mathrm{sgn}(\frac{\sin k}{\partial \mathbf{k} }\mid_{\mathbf{K_i}})\\
  &\quad \mathrm{sgn}((t^2+1+2t\cos k-\delta^2-\mu^2)\mid_{\mathbf{K_i}}).
  \end{split}
\end{equation}
with $\mathbf{K_i}$ is the $i-$th solution of $\sin k=0$. This is different from the Hermitian cases where $\mathbf{K_i}$ is determined by $\hat{h}_0=0$.

Now we give the geometric meaning of the winding number $\nu_E$. To see this, we parameterize the square of energies by:
\begin{eqnarray*}
   E^2_{1,2} &=& |E|^2e^{2i\theta_k},
\end{eqnarray*}
with
\begin{eqnarray*}
   \tan 2\theta_k &=& \frac{2 \delta \sin k}{t^2+1+2t\cos k-\delta^2-\mu^2}=\frac{\mathrm{Im}(E_{1,2}^2)}{\mathrm{Re}(E_{1,2}^2)}.
\end{eqnarray*}
then the winding number can be written as
\begin{equation*}
\begin{split}
2\pi\nu_E=&\oint dk \nabla_k \theta_k= \oint dk\frac{1}{2} \cos 2\theta \nabla_k\tan 2\theta \\
   &=\frac{2\mathrm{Re}(E_{1,2}^2)}{|E_{1,2}^2|^2} \nabla_k\frac{\mathrm{\mathrm{Im}}(E_{1,2}^2)}{\mathrm{Re}(E_{1,2}^2)}\\
&=\oint dk\frac{\mathrm{Re}(E_{1,2}^2)\nabla_k \mathrm{Im}(E_{1,2}^2)-\mathrm{Im}(E_{1,2}^2)\nabla_k \mathrm{Re}(E_{1,2}^2)}{2|E_{1,2}^2|^2}\\
&=\oint dk\frac{\mathrm{Re}(E_{1,2}^2)\nabla_k \alpha \mathrm{Im}(E_{1,2}^2)-\alpha \mathrm{Im}(E_{1,2}^2)\nabla_k \mathrm{Re}(E_{1,2}^2)}{2(\mathrm{Re}(E_{1,2}^2)^2+\alpha^2\mathrm{Im}(E_{1,2}^2)^2)}\\
   & =\mathrm{sgn}(\alpha)\oint dk \nabla_k \theta',
   \end{split}
\end{equation*}
where $\tan 2\theta'=\frac{\alpha \mathrm{Im}(E_{1,2}^2)}{\mathrm{Re}(E_{1,2}^2)}$. Here $\alpha$ is independent of $k$ and taken to be $\alpha =\frac{\sqrt{\mu^2+\delta^2}}{\delta}$, thus $\mathrm{sgn}(\alpha)=\mathrm{sgn}(\delta)$. The winding number of $\theta_k$ is now represented by the winding number of $\theta'_k$ as shown in Fig. 7(a). Furthermore, we have
\begin{eqnarray*}
  \mathrm{Re}(E_{1,2}^2)+i\alpha \mathrm{Im}(E_{1,2}^2) &=& \sqrt{(\mathrm{Re}(E_{1,2}^2)^2+\alpha^2\mathrm{Im}(E_{1,2}^2)^2)}e^{2i\theta'}\\
   &=& \sqrt{(\mathrm{Re}(E_{1,2}^2)^2+\alpha^2\mathrm{Im}(E_{1,2}^2)^2)}e^{-i\phi_1}e^{i\phi_2},
\end{eqnarray*}
with $\tan \phi_1=\frac{n_y}{n_x+\sqrt{\mu^2+\delta^2}}$ and $\tan \phi_2=\frac{n_y}{n_x-\sqrt{\mu^2+\delta^2}}$, where $n_x=t+\cos k$ and $n_y=\sin k$. Here $\phi_1$ and $\phi_2$ are the angles of vector $\mathbf{n}(k)$ around the two EP points $EP_1$ and $EP_2$ as shown in Fig.7(b), respectively. Finally, the winding number $\nu_E$ becomes
\begin{equation}\label{gg}
\nu_E=\mathrm{sgn}(\delta)\frac{1}{2\pi}\oint dk \nabla_k \theta'=\frac{1}{2}\mathrm{sgn}(\delta)({\nu}_2-{\nu}_1),
\end{equation}
where $\nu_i=\frac{1}{2\pi}\oint dk \nabla_k \phi_i$. Hence $\nu_E$ measures the differences of winding number ${\nu}_1$ and ${\nu}_2$, which is similar to the case of chiral Hamiltonian discussed in Ref.\cite{Yin}.

\begin{figure}[h]
\centering
\includegraphics[width=1\columnwidth]{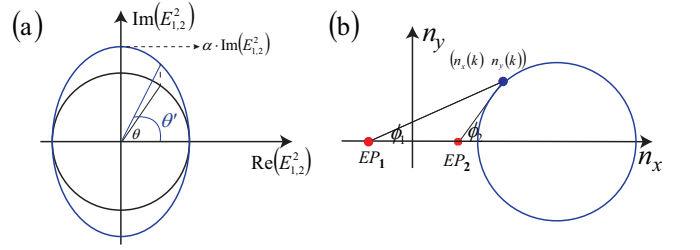}
\caption{(a) A schematic diagram of  compressive deformation  $E_{1,2}^2$.  (b) A schematic diagram shows the geometrical meaning of $\phi_1$ and $\phi_2$ with $n_{x/y}$=Re $\langle\sigma_{x/y}\rangle$}.
\end{figure}\label{SFig1}

\section{The winding of eigenstate $\nu_s$}
The eigenstates for non-Hermitian Hamiltonian satisfy
\begin{eqnarray*}
 H(\mathbf{k})|\psi^{R}_{1,2}\rangle&=E_{1,2}|\psi^{R}_{1,2}\rangle, \\ \langle\psi^{L}_{1,2}|H^{\dagger}(\mathbf{k})&=\langle\psi^{L}_{1,2}|E_{1,2},
\end{eqnarray*}
with
\begin{eqnarray*}
|\psi^{R}_{1,2}\rangle &=\frac{1}{\sqrt{2E_{1,2}(E_{1,2}-h_z)}}\left(
    \begin{array}{cc}
        h_x-ih_y,E_{1,2}-h_z
    \end{array}
\right)^{T},\\
\langle\psi^{L}_{1,2}|&=\frac{1}{\sqrt{2E_{1,2}(E_{1,2}-h_z)}}\left(
    \begin{array}{cc}
        h_x+ih_y,E_{1,2}-h_z
    \end{array}
\right),
\end{eqnarray*}
where the superscript $T$ is transpose operation. The Berry phase $\nu_s$ of the state is defined by
\begin{equation*}
  \nu_{s,1}=\frac{1}{\pi}\oint dk \langle\psi^{L}_{1}|i\partial_{k}|\psi^{R}_{1}\rangle.
\end{equation*}
Substituting the expression of $|\psi^{R}_{1}\rangle$ ,$\langle\psi^{L}_{1}|$ into this equation, $\nu_s$ is rewritten as
\begin{equation*}
 \begin{split}
  \nu_{s,1}  &= \frac{1}{\pi}\oint dk \frac{1}{\sqrt{2E_{1}(E_{1}-h_z)}}\left(\begin{array}{cc}
                                                                                   h_x+ih_y &E_1-h_z \\
                                                                                 \end{array}
                                                                               \right)  \\
                                                                             & \qquad \qquad \qquad \qquad i\partial_{k}\frac{1}{\sqrt{2E_1(E_1-h_z)}}\left(
                           \begin{array}{cc}
                             h_x-ih_y \\E_1-h_z \\
                           \end{array}
                         \right)\\
            &=\frac{1}{\pi}\oint dk \frac{h_x\partial_{k}h_y-h_y\partial_{k}h_x}{2E_1(E_1-h_z)},\\
 \end{split}
\end{equation*}
Summing up the Berry phases of the two bands, the total Berry phase is
\begin{eqnarray*}
  \nonumber\nu_{\mathrm{tot}} = \nu_{s,1}+\nu_{s,2}=  \frac{1}{\pi}\oint dk \frac{h_x\partial_{k}h_y-h_y\partial_{k}h_x}{h_x^2+h_y^2},
\end{eqnarray*} \label{vstotal}
which can be proved to be quantized.

In a Hermitian system, a Hamiltonian having inversion symmetry means there is an unitary operator satisfying $UH(k)U^{-1}=H(-k)$. As a comparison, we can define a pseudo-inversion symmetry in the non-Hermitian system. Because of  $H(k)\neq H^{\dagger }(k)$, the pseudo-inversion symmetry now requires $UH(k)U^{-1}=H^{\dagger}(-k)$, while the operator $U$ is still unitary. For example, if $U$ is chosen to be $\sigma_x$, the pseudo-inversion symmetry gives some constrains on the Hamiltonian, i.e.,
\begin{eqnarray*}
     h_{x}(k) &=&h^{*}_{x}(-k), \\
     h_{y}(k) &=& -h^{*}_{y}(-k), \\
     h_{z}(k) &=& -h^{*}_{z}(-k).
\end{eqnarray*}
Besides, the eigenvalues should satisfy $E_1(k)=E^{*}_1(-k)$, or $E_1(k)=E^{*}_2(-k)$. Now we study the Berry phase for these two cases, respectively.

In the first case, we have $E_1(k)=E^{*}_1(-k)$, and the Berry phase $\nu_{s,1}$ is
\begin{eqnarray*}
\begin{split}
&\nu_{s,1}=\frac{1}{\pi}\int^{\pi}_{-\pi} \mathrm{d} k \langle\psi^{L}_{1}|i\partial_{k}|\psi^{R}_{1}\rangle\\
         &=\frac{1}{\pi}\int^{\pi}_{-\pi} \mathrm{d} k \mathrm{d}k \frac{h_x(k)\partial_{k}h_y(k)-h_y(k)\partial_{k}h_x(k)}{2E_1(k)(E_1(k)-h_z(k))}\\
         &=\frac{1}{\pi}\int^{\pi}_{-\pi} \mathrm{d} k \frac{-h^*_x(-k)\partial_{k}h^*_y(-k)+h^*_y(-k)\partial_{k}h^*_x(-k)}{2E^*_1(-k)(E^*_1(-k)+h^*_z(-k))}\\
         &=\frac{1}{\pi}\int^{\pi}_{-\pi} \mathrm{d}(-k) \frac{h^*_x(-k)\partial_{-k}h^*_y(-k)-h^*_y(-k)\partial_{-k}h^*_x(-k)}{2E^*_1(-k)(E^*_1(-k)+h^*_z(-k))}\quad (k\rightarrow-k)\\
          &=\frac{1}{\pi}\int^{\pi}_{-\pi} \mathrm{d}k \frac{h^*_x(k)\partial_{k}h^*_y(k)-h^*_y(k)\partial_{k}h^*_x(k)}{2E^*_2(k)(E^*_2(k)-h^*_z(k))}\\
          &=\frac{1}{\pi}\int^{\pi}_{-\pi} \mathrm{d} k \langle\psi^{R}_{2}|i\partial_{k}|\psi^{L}_{2}\rangle\\
          &=\nu_{s,2}^*.
   \end{split}
\end{eqnarray*}
Similarity, we can see $\nu_{s,2}=\nu_{s,1}^*$. As a result, the total Berry phase $\nu_{tot}=\nu_{s,1}+\nu_{s,2}=\nu_{s,1}+\nu_{s,1}^*$ is real and quantized. The real and imaginary part of $\nu_{s,1}$ satisfy $\mathrm{Re}(\nu_{s,1})=\mathrm{Re}(\nu_{s,2})= \frac{1}{2}\mathrm{Re}(\nu_{tot})$; $\mathrm{Im}(\nu_{s,1})=\mathrm{Im}(\nu_{s,2}^{*})$. This phase is called pseudo-inversion symmetry unbroken phase, in which the real part of Berry phase $\nu_{s,i}$ is quantized.

In the second case, $E_1(k)=E^{*}_2(-k)$. The Berry phase $\nu_{s,1}$ is
\begin{eqnarray*}
\begin{split}
&\nu_{s,1}=\frac{1}{\pi}\int^{\pi}_{-\pi} \mathrm{d} k \langle\psi^{L}_{1}|i\partial_{k}|\psi^{R}_{1}\rangle\\
         &=\frac{1}{\pi}\int^{\pi}_{-\pi} \mathrm{d} k \mathrm{d}k \frac{h_x(k)\partial_{k}h_y(k)-h_y(k)\partial_{k}h_x(k)}{2E_1(k)(E_1(k)-h_z(k))}\\
         &=\frac{1}{\pi}\int^{\pi}_{-\pi} \mathrm{d} k \frac{-h^*_x(-k)\partial_{k}h^*_y(-k)+h^*_y(-k)\partial_{k}h^*_x(-k)}{2E^*_2(-k)(E^*_2(-k)+h^*_z(-k))}\\
          &=\frac{1}{\pi}\int^{\pi}_{-\pi} \mathrm{d}k \frac{h^*_x(k)\partial_{k}h^*_y(k)-h^*_y(k)\partial_{k}h^*_x(k)}{2E^*_1(k)(E^*_1(k)-h^*_z(k))}\\
          &=\frac{1}{\pi}\int^{\pi}_{-\pi} \mathrm{d} k \langle\psi^{R}_{1}|i\partial_{k}|\psi^{L}_{1}\rangle\\
          &=\nu_{s,1}^*.
   \end{split}
\end{eqnarray*}
In this case, the $\nu_{s,1}$ and $\nu_{s,2}$ are real but not quantized. The phase is called pseudo-inversion symmetry broken phase.



\end{document}